Considering Risk Aversion in Economic Evaluation: A Rank Dependent Approach.

Jacob Smith*, University of Toronto IHPME
155 College St 4th Floor, Toronto, ON M5T 3M6
ORCID: https://orcid.org/0000-0001-7179-9567

Abstract

This paper presents a method for incorporating risk aversion into existing decision tree models used in economic evaluations. The method involves applying a probability weighting function based on rank dependent utility theory to reduced lotteries in the decision tree model. This adaptation embodies the fact that different decision makers can observe the same decision tree model structure but come to different conclusions about the optimal treatment. The proposed solution to this problem is to compensate risk-averse decision makers to use the efficient technology that they are reluctant to adopt.

Introduction

The purpose of economic evaluation is to inform the decision to invest (or not invest) in a novel health technology from the perspective of risk neutrality. Risk neutrality is an important assumption as it has much to do with the provision of goods and services for a large population with a uniform distribution of health states. However, there is recent work which demonstrates that medical decision makers are in fact risk averse (see (Arrieta, García-Prado, González, & Pinto-Prades, 2017) and (Saposnik, Redelmeier, Ruff, & Tobler, 2016)).  With these empirical facts we ask and attempt to answer the following questions:

1) How are medical decisions impacted by risk aversion and how does risk aversion manifest itself in clinical environments?
2) How do we adapt pre-existing decision tree models to account for risk aversion?
3) What useful statistics can be generated from decision analytic models with risk averse measures within them?

The objective of this paper is one which will attempt to provide an applied framework for accounting for and adapting existing decision tree models used in cost effectiveness analysis to account for risk aversion.

How Risk Aversion in Medical Decision Making is Observed

What does risk aversion look like in the context of medical decision making? In simple review of the broader health services research literature, we find several ways in which the term "risk averse" is used to describe medical decision-making behaviour. The main three ways we have seen this term used are:

● Risk aversion in medical decision-making means persistence in the use of standard treatments considering better treatment options (Saposnik, Redelmeier, Ruff, & Tobler, 2016), (Ramani, Gilson, Sivakami, & Gawde, 2021).
● Risk aversion in medical decision-making means excess of testing (Michel-Lepage, Ventelou, Nebout, Verger, & Pulcini, 2013).

● Risk aversion in medical decision-making means either the delay or speed up of time to treatment. (Cykert, 2004), (Shahian, et al., 2017)

Though these are all medically relevant results of interests to clinicians, for those who allocate hospital budgets (such as purchasing managers) what does this mean? Firstly, it means that optimal investment in new health technology by managers does not lead to optimal health outcomes and realized net benefits from that investment. With risk aversion as seen in the persistence of standard care when better technology is available because of investment in it, decision makers incur all the costs associated with investment in new technology and miss out on the health benefits that could be realized. Secondly (with reference to the latter two working definitions of risk aversion in health) we have a case where the model used by health policy analysts is not the same as that used by clinicians. This argument has roots with the of (Asch & Hershey, 1995) which emphasises that clinicians do not make decisions with reference to population health outcomes, but rather focus on outcomes of individuals. Analytically this means that the model used in defining if a new technology is cost effective will be modified through the addition of new testing and a discounting of health benefits to be realized much later than intended.

Optimal investment in health technology requires a harmony between the choice of hospital purchasing managers and the choices of clinicians. If there is a discrepancy between the choices of these two agents, we have a case of non optimal investment. To account for this discrepancy, we note that we should be explicitly clear in how the same information which is built into our decision analytic models are handled by clinicians. To begin this conversation, we will start with a review of decision tree models and how they are used in the economic evaluation of health technologies.

Adapting Decision Tree Models to cases of Risk Aversion

The most common model seen in economic evaluation are decision tree models. The reason why they are so popular in health technology assessment is because they represent the view of clinicians as they see possible treatment options. They are visually appealing because they are visualizations of the choice over expected net benefits. Analytically the model we are solving in for decision-tree based health technology assessment models is:

$$\arg\max_{j \in \mathcal{J}} \left\{ \sum_{s \in \mathcal{S}} p(s)\, NB(j, s) \right\}$$

Where:

$$NB(j, s) = \lambda E_j(s) - C_j(s)$$

In this framework we are seeking the technology $j$ from a list of technologies $\mathcal{J}$ which yields the highest expected net benefit considering all possible states of the world $s$ in $\mathcal{S}$. In this framework the expected net benefit of each technology $j \in \mathcal{J}$ serves as an index for the optimal choice of technology. It is this framework which defines the type of choices to be made considering various willingness to pay thresholds $\lambda$.

Further it is noted that in this framework we are defining the decision model as one which makes the optimal choice of health technology from the perspective of risk neutrality. Health indexes such as the EQ-5D-5L measures health states without consideration for risk aversion, so how would one modify this framework to account for such behaviour? One simple solution would be to transform our health utilities index under $E_j(s)$ for each $j \in \mathcal{J}$ and $s \in \mathcal{S}$ through the use of a function (a monotone transformation). This solution is problematic because it violates the interval property and can even violate condition 0 associated with the utility index (this leaves dead states undefined if our transformation is logarithmic). A less problematic approach to making these considerations would be to focus instead on a transformation of the probabilities with the use of the probability weighting function in a manner that is in line with rank dependent utility theory (Quiggin, 1982) or more famously seen in prospect theory (Kahneman & Tversky, 1979). Our ability to move to this framework conceptually has been made easy by (Sarin & Wakker, 1998) as will be expounded upon in the next section.

Rank dependent utility and its hurdles to use in health technology assessment.

As previously stated, our approach to describing our medical decision maker's view of uncertainty and in turn characterize risk aversion is by using the framework of rank dependent utility. To clarify our objective problem of our medical decision maker becomes:

$$\arg\max_{j \in \mathcal{J}} \left\{ \sum_{s \in \mathcal{S}} \pi(p(s)) NB(j, s) \right\}$$

Where:

$$NB(j, s) = \lambda E_j(s) - C_j(s)$$

$$\pi(p(s)) = \frac{p(s)^\gamma}{\left(p(s)^\gamma + (1 - p(s))^\gamma\right)^{\frac{1}{\gamma}}}$$

Notice how we have not touched our net benefit equations at all but have instead transformed our probabilities through the standard probability weighting given in (Kahneman & Tversky, 1979)[1].

Note that before we can go on our way with this modified framework for decision analysis in health technology assessment, we must note that there are some additional assumptions which must be made of our model. First, we note that one of the main tools we use for solving our decision trees is through "folding back" by representing the choice over uncertain outcomes to one which is a choice over means. To illustrate consider the case where we have a clinician choosing between two treatments, treatment A and B with two layers of unique risk associated with each treatment line. We can visualize this model as that in Figure 1.

---

[1] Transforming the net benefits associated with each treatment $j$ in state $s$ through a value function similar to a prospect theoretic model would abstract too far away in terms of the meaning we seek to derive from this (i.e. Net benefits is a dollar value, what is a transformed dollar?). This isn't to say that there isn't a diminishing value from net benefits according to our decision maker's criteria and subsequently a risk aversion associated with gambles across net benefits or in the prospect theoretic sense risk seeking behaviour among losses. It just becomes increasingly difficult to bring back to a specific policy variable.

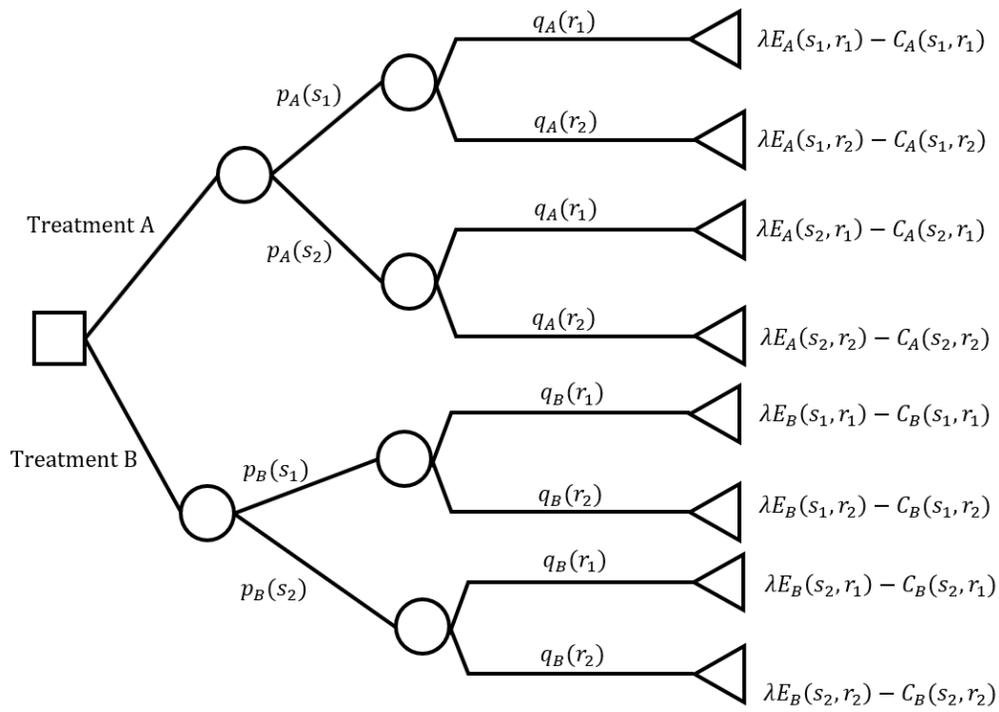

*Figure 1: Modelling a choice between two treatments in the context of a basic decision tree with two layers of uncertainty.*

An equivalent set of choices can be modelled by considering the choice over reduced lotteries in Figure 2. This choice can be further pushed back to the choices over expectations, looking at single statistics. We can think that the modelling of choice under uncertain outcomes is one which transforms objects which we cannot take hold of just yet (such as the unrealized outcome in question) and gives us a statistic which decision makers can hold onto known as the expectation. Though this result is just a restatement of the obvious we ought to ask what are the mechanisms which allow us to move smoothly between these representations of choice between theses certainty equivalents and choices between lotteries over uncertain outcomes.

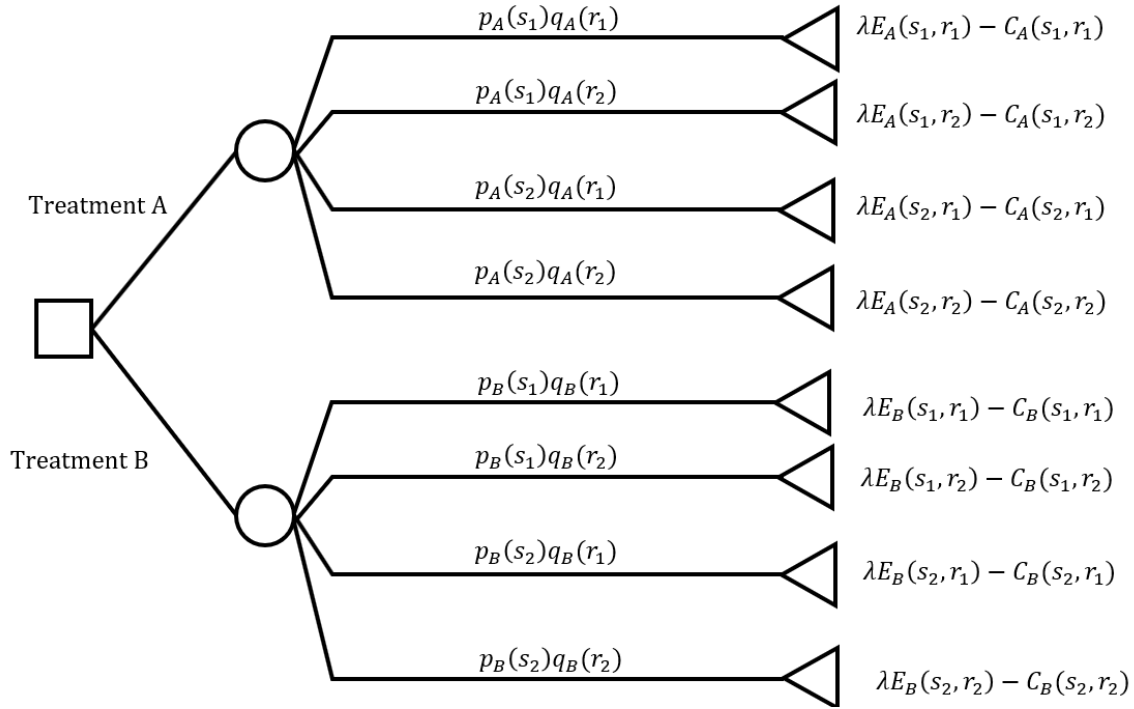

*Figure 2: A reduced form representation of our problem in Figure 1.*

The main device which we can use is the law of total probability. It is this rule which assigns a reason to how our certainty equivalents come into existence and keeps our probability space summing to one always no matter how many layers of uncertainty a problem may have.

The rank dependent framework sacrifices the notion of certainty equivalent in addition to the law of total probability. Though we can use this framework to rank treatment options, the statistics that we produce from it are obscure to the applied researcher. Further it is not quite clear how our probability weighting function ought to be applied. To illustrate, we ask: Is it more (a) appropriate to apply our probability weighting function at each chance node in our model or (b) must we consider the reduced form model where our probabilities compound? The difference between these two options can be seen in Figures 3 and 4 respectively.

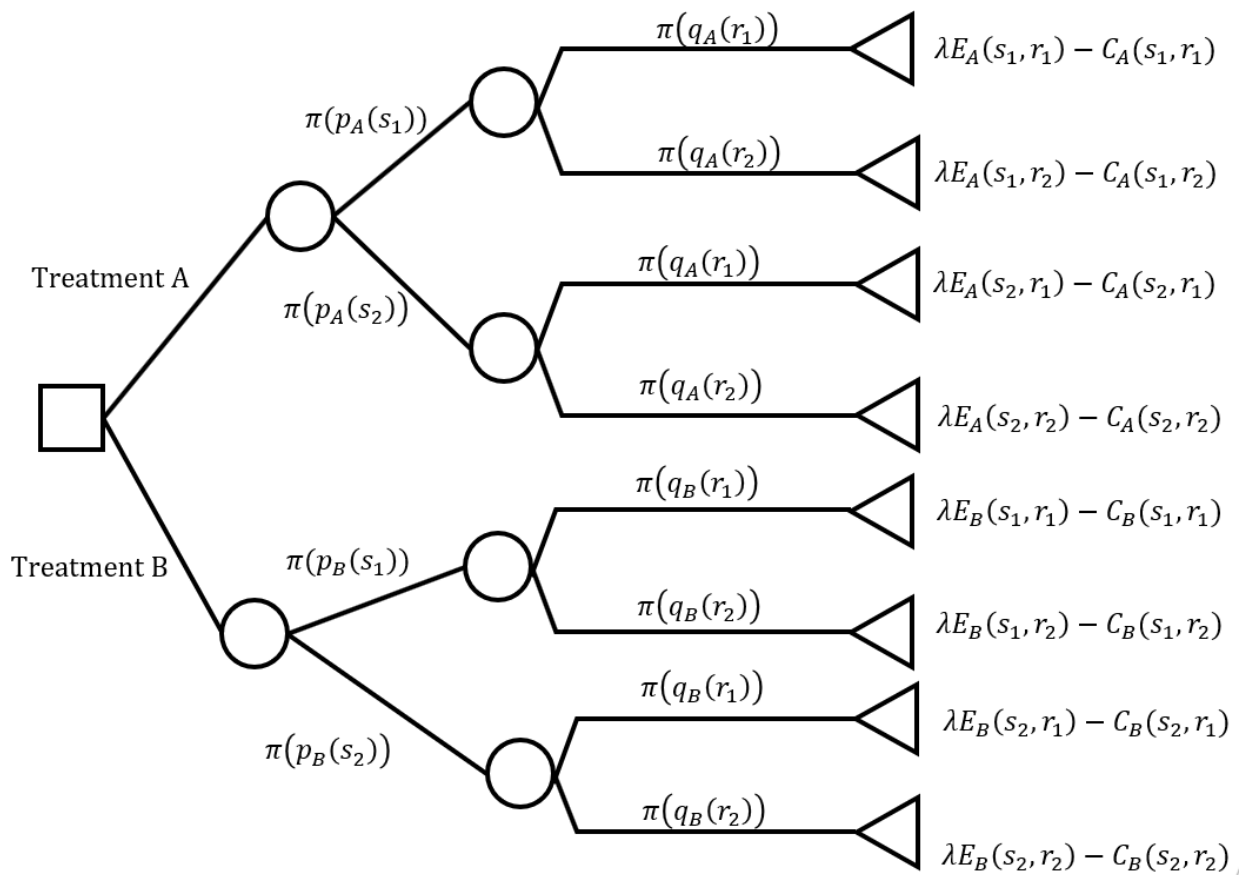

Figure 3: Applying the Probability weighting function at each chance node.

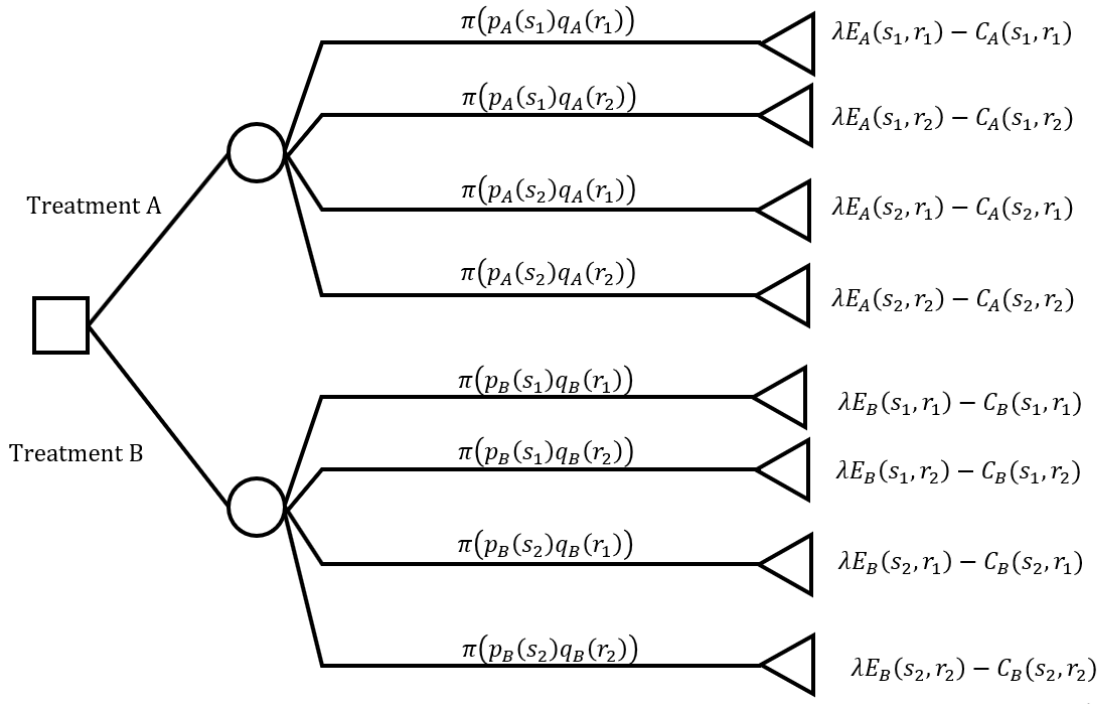

*Figure 4: Applying the probability weighting function over reduced lotteries.*

The rule to follow in terms of a computational framework is to pick either one of these two frameworks but never a combination of these two. The way in which this is informed by results are due to (Sarin & Wakker, 1998) in which we utilize the condition of "sequential consistency". Simply put, define how our decision maker applies the probability weighting function.

For our purposes we will be assuming that the probability weighting function will be applied as it is over reduced lotteries. The reason why this is done is because of the difficulty of interpretation that multiplication of each weight goes and has. Through this probability weighting function that we can see the difference in treatment choice, an example of this switching example can be seen in the appendix where we compare the model output from a (slightly modified) sample example presented in (Drummond, Sculpher, Torrance, O'Brien, & Stoddart, 2005).

Operationalizing the Rank Dependent Model through the lens of Value of Information Analysis

How do we use our rank dependent valuation of our decision tree model in applied work of health technology assessment? The purpose of economic evaluation of health technologies is to define a discrete decision of using one treatment $j \in \mathcal{J}$ over all others conditional on a specific willingness-to-pay. Its clear that there can be a difference between the decision of the optimal technology under our standard model compared to the rank dependent framework, but how can we possibly visualize these differences. One option is to make a comparison in terms of the expected value of perfect information (EVPI) and the rank dependent version of this statistic which we call "Rank dependent value of perfect information" (RDVPI). The way both these statistics are calculated is through the use of the following equations:

$$EVPI(\lambda) = \sum_{s \in S} p(s) \max_{j \in J}\{NB(j,s)\} - \max_{j \in J}\left\{\sum_{s \in S} p(s) NB(j,s)\right\}$$

$$RDVPI(\lambda) = \sum_{s \in S} p(s) \max_{j \in J}\{NB(j,s)\} - \max_{j \in J}\left\{\sum_{s \in S} \pi(p(s)) NB(j,s)\right\}$$

In the usual fashion of calculating the value of perfect information our willingness to pay threshold is varied and repeated computation of these value of information equations are generated. A characteristic of this computing is the fact that there can be a kink in our value of information curve which will be due to a switch in optimal treatment due to changes in willingness to pay threshold $\lambda$. The exact way in which these switches occur is through the shift in strategy in the second term on the LHS with the "switch point threshold" for both the EVPI curve and RDVPI curve being denoted as $\hat{\lambda}_{EV}$ and $\hat{\lambda}_{RD}$ respectively. An illustration of the two curves generated appears in the figure below.

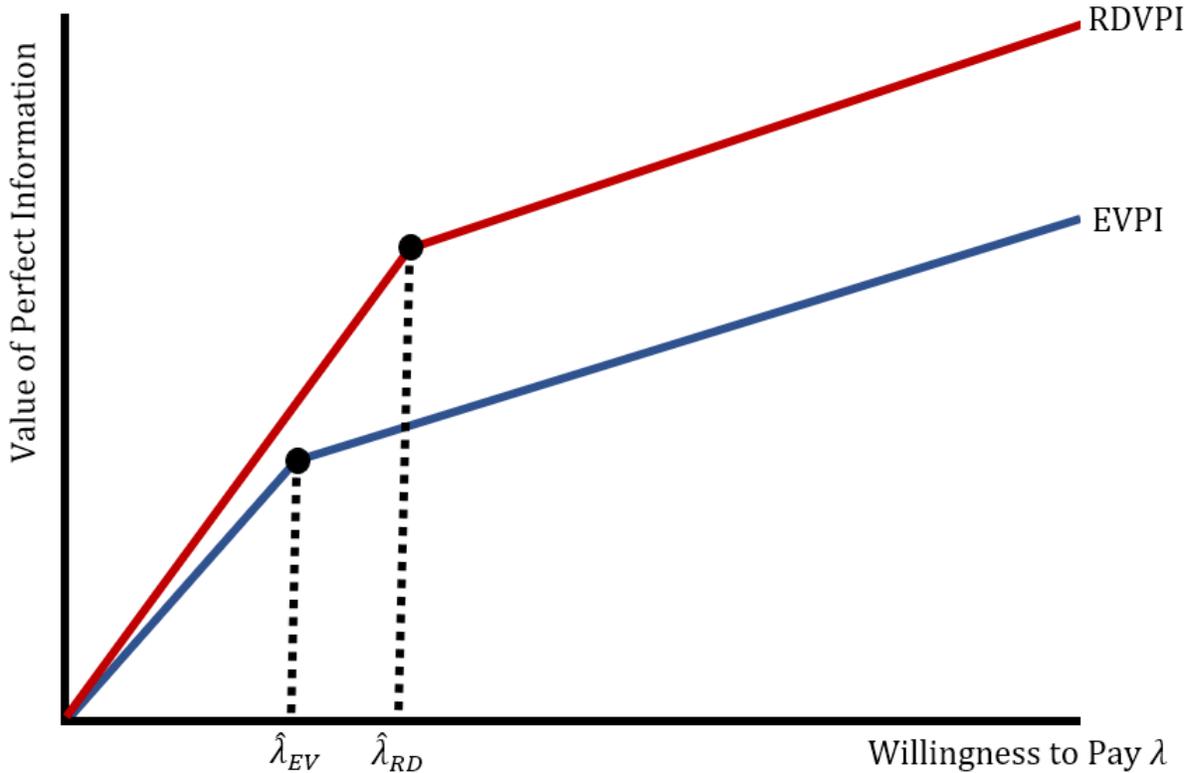

*Figure 5: Comparing Switch Points*

Before we move forward, we should take some time to appreciate what is being communicated in this diagram. Firstly, there is an assumption that $RDVPI(\lambda) \geq EVPI(\lambda)$ for all values of $\lambda$. From an outset, this is a problematic assumption if our focus is on the exact comparison between these two statistics because of existing work by (Eekhoudt, 2002) which shows that depending on the nature of risk and preferences in our models the sign in this relationship can change. We restrict our attention to this case because it is the type of risk aversion which results in non optimal treatment. To see why such a result occurs, let's consider four possible graphs of our VPI curves as seen in Figure 6.

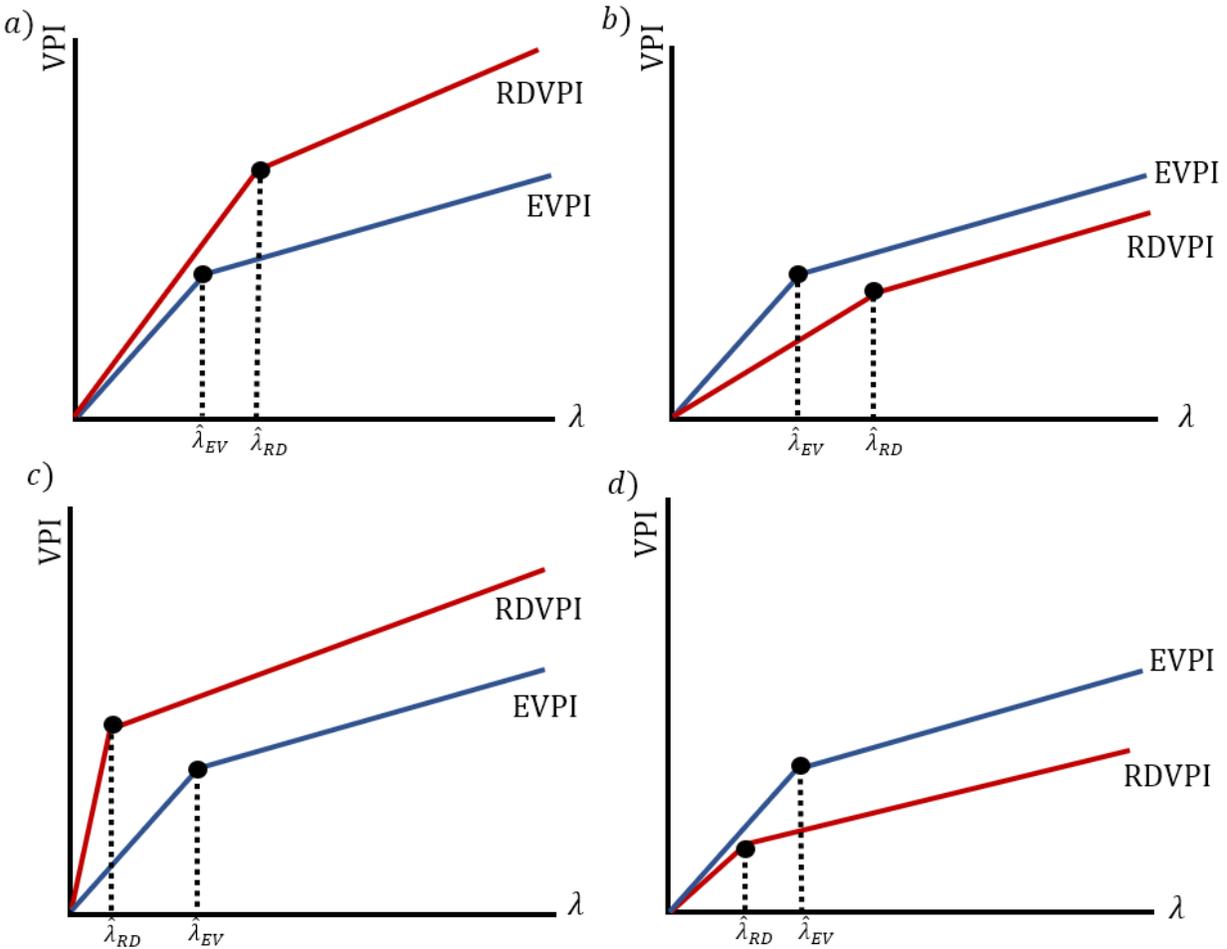

Figure 6: Four Possible Scenarios

In this diagram we see a difference both in the magnitude of the value of information curves and a difference in the possible switch points. We can note right away in panels c) and d) that this is not the type of risk aversion we are interested in. This is because we have a case where our risk averse medical decision maker would adopt an alternative at a lower threshold compared to our risk neutral decision maker. Even though their values differ in terms of the ordering of $EVPI(\lambda)$ with respect to $RDVPI(\lambda)$ there is coordination in the context of optimal treatment because $\hat{\lambda}_{RD} < \hat{\lambda}_{EV}$.

This leaves us with two possible scenarios of $EVPI(\lambda)$ and $RDVPI(\lambda)$, panels a) and b) where our risk averse medical decision maker is adopting an alternative technology at a higher threshold. Looking at panel b) we can see that this is a scenario where the value of information added to our risk averse decision maker is lower than that of our risk neutral one. We can imagine this as the case where we have refusal to switch and a refusal to acquire more information to treat despite our risk neutral decision maker seeing value. The implication of this graph is that we have risk aversion which is not described by excess testing but is described by the persistence in treatment. In panel a) however we have an image which fits both a description of risk aversion as a use of excess testing and persistence in use of standard care as seen by $\hat{\lambda}_{RD} > \hat{\lambda}_{EV}$ and $RDVPI(\lambda) > EVPI(\lambda)$ for all willingness-to-pay thresholds $\lambda$.

Harmonizing the Decisions of Risk Averse and Risk Neutral Decision makers

As seen in the above section the way we model the lack of coordination between our risk neutral and risk averse decision makers is through the observation that the ordering of switching points differ due to $\hat{\lambda}_{RD} > \hat{\lambda}_{EV}$. We can calculate the difference between these two switching thresholds (from here on referred to as the compensating threshold) as:
$$\varepsilon = \hat{\lambda}_{RD} - \hat{\lambda}_{EV}$$
Thus to harmonize the choices of optimal treatments, to that of the risk neutral decision maker's problem, we require that for the alternative treatment $\hat{j}$ it must yield a net benefit of:
$$NB(\hat{j}, s) = (\lambda + \varepsilon) E_j(s) - C_j(s)$$

Or
$$NB(\hat{j}, s) = \lambda E_j(s) - C_j(s) + \varepsilon E_j(s)$$

What this means is that for the optimal treatment by our risk neutral decision maker to correspond to the optimal treatment according to our risk averse decision maker we must compensate the net benefits of the alternative. The primary policy device for using this is by first noting that $\varepsilon$ is a monetary amount, as such the strategy for optimal implementation of a new technology which clinicians are risk averse to should be oriented around proper compensation for its use. To obtain optimal implementation we ought to compensate clinicians for their risk-taking in monetary terms according to this modified net benefits equation. This amount is defined by $\varepsilon E_j(s)$ for each health state which can be realized by technology $\hat{j}$. Given that we are using a rank dependent valuation this monetary value is $\sum_{s \in S} \pi(s) \varepsilon E_j(s)$.

Limitations of this approach

In this framework we have pointed to the key policy variable we can use to align the decisions of analysts and clinicians is through direct compensation in the amount of our "rank dependent valued effects" and our compensating threshold $\varepsilon$. This approach though can result in "over payment" of clinicians due to the magnitude of $\varepsilon$ which is due to the size of the intervals between thresholds $\lambda$. Further the way effects are measured can influence the compensating payment dramatically (i.e., if we are using QALYs which are bounded on the interval of 0 and 1 this amount would be smaller compared to some positive index with no such rule).

Future work should uncover the dynamics in the southwest cost-effectiveness plane in which cost-effective interventions are by necessity less effective than standard interventions albeit at less cost. In such instances, providers may be risk-loving rather thank risk-averse alongside exhibiting loss-aversion according to Prospect Theory (Kahneman and Tversky, 1979 and Tversky and Kahneman, 1992)

## Conclusion

In this paper we explored how we can both frame and adapt existing decision tree models for the economic evaluation of health technologies using a rank dependent framework. We do this by first discussing why one would want to consider risk aversion all together in economic evaluation and motivate this by the asymmetries that exist between managers investing in new health technologies and the use of those technologies by clinicians. The solution we propose to this problem is through compensating clinicians for the use of the alternative technology according to the difference of switching thresholds and effects valued by the probability weighting function in each state.

Appendix: An example of switching in optimal treatment from model change.

To demonstrate the differences in valuation using expected values vs valuation through the use of a rank dependent valuation, we note a simple (slightly modified) example decision tree model from (Drummond, Sculpher, Claxton, Stoddart, & Torrance, 2015). In this case we consider the comparison of a surgical intervention to a baseline medical one for some arbitrary condition. A display of this model is seen in Figure 7.

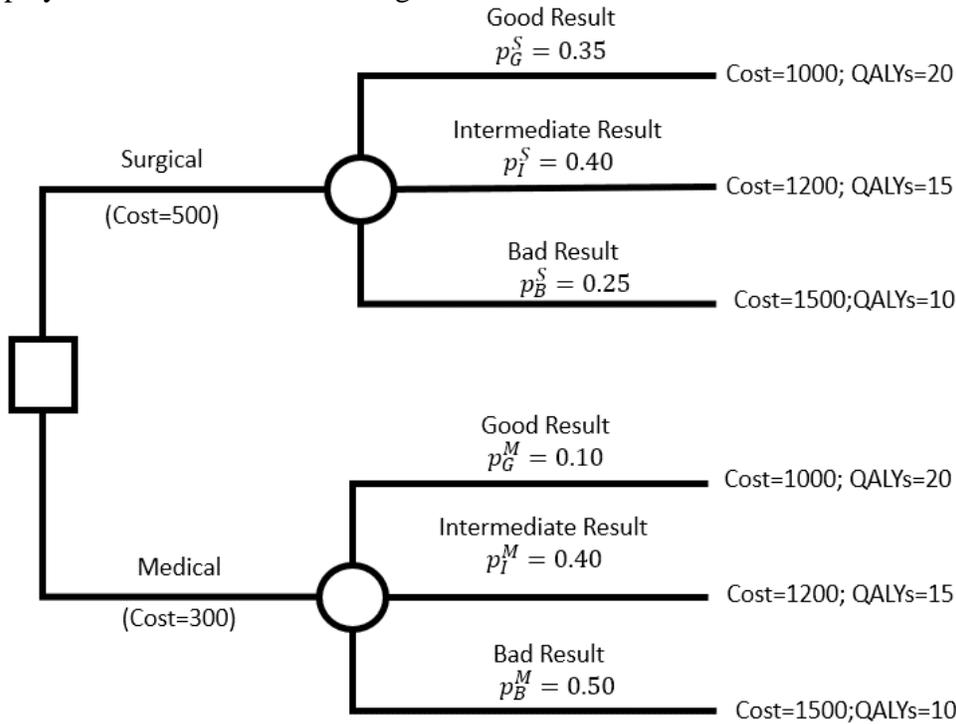

*Figure 7: A basic decision tree model.*

The orientation of this decision tree in its current state makes it difficult to employ the valuations given by prospect theory. Prospect theory at its core is a theory of choice over ranked outcomes in states of nature, in our current orientation we are considering two distinct outcomes- Costs Borne and QALYs gained. To address this issue of "simultaneous outcome" we define payoffs on the end of each node as a **Net Monetary Benefit (NMB)** where each terminal value is written as:

$$NMB = E_i^d \lambda - C_i^d$$

Where:

- $E_i^d$ is the effectiveness of the choice of treatment $d$ in state $i$.
- $C_i^d$ is the cost of the choice of treatment $d$ in state $i$.
- $\lambda$ a threshold value.

This orients our decision tree to be as follows:

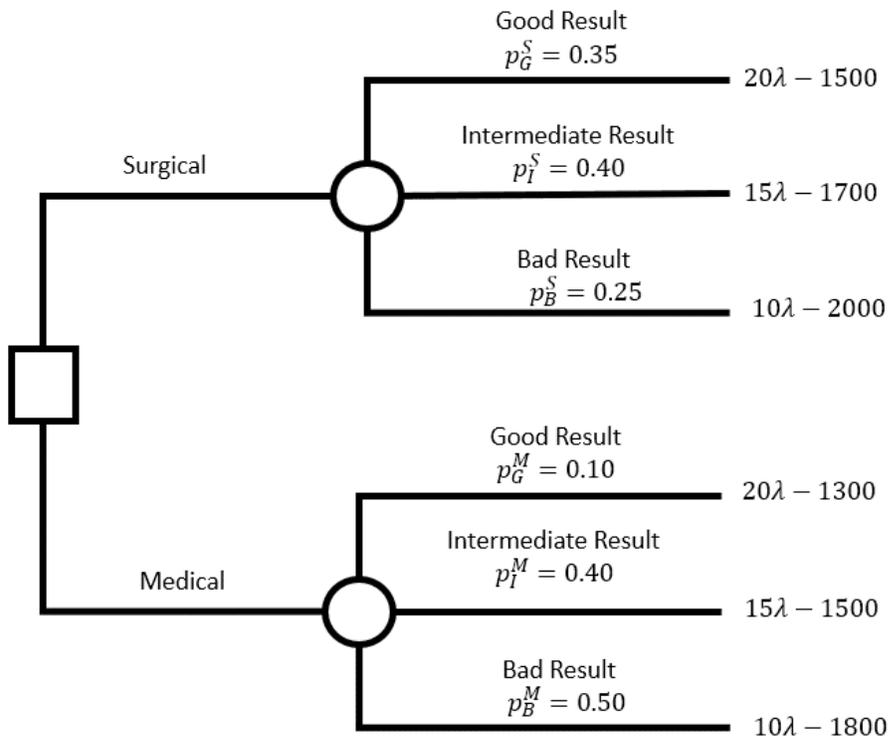

*Figure 8: A linearized version of the decision tree model.*

This framework further gives us the ability to simply define which treatment is more cost effective by just comparing their valuation whether that be with the use of conventional expected values or in the context of prospect theory.

One thing should be noted before we compare our valuations. The only function from prospect theory which we use is the probability weighting function. We do not transform our net benefits through the value function. This is done for the fact that we are already considering a transformed value (costs and QALYs to net monetary benefits). Transforming these values further would only muddle these results. The probabilities which we are given are fit for our weighting functions. A comparison of the prospect theoretic valuation of each arm of the decision tree with comparison to the corresponding baseline expected value is seen in the table below.

|  | Prob weights Surgical | Prob Weights Med | Parameters | Parameter Value |
|---|---|---|---|---|
| $p_G$ | 0.09958885 | 0.100935727 | $\gamma$ | 0.25 |
| $p_I$ | 0.02977579 | 0.004176325 | $\lambda$ | 1000 |
| $p_B$ | 0.87063536 | 0.894887948 |  |  |
| SUM | 1.00 | 1.00 |  |  |
|  |  |  |  |  |
| Choice | Rank Dependent Valuation | Expected Value |  |  |
| Surgical | 9203.495 | 13795✓ |  |  |
| Medical | 9281.96✓ | 11370 |  |  |
|  |  |  |  |  |

*Table 1: A Simple example of switched optimums from change in model (from EV to RDV).*

The most important thing to note from this example is that the choice of best option changes from the surgical option to the medical when employing our probability weighting function in these values. The significance of these results is that the way uncertainty is assessed can change the individual valuation each option.

In relating these results back to our proposed solution for dealing with physician risk aversion we consider a small interval of thresholds from [0,5000] and note the optimal response according to each model by threshold and simply note the difference between thresholds driving the change in response. Sample output can be seen in Table 2.

| Threshold | Surgical RD | Medical RD | Optimal RD Choice | Surgical EV | Medical EV | Optimal EV Choice | Compensating Threshold $\varepsilon$ |
|---|---|---|---|---|---|---|---|
| 0 | -1941.27 | -1748.28 | Medical | -1205 | -1330 | Surgical | 0 |
| 1000 | 9203.495 | 9281.96 | Medical | 345 | -30 | Surgical | 0 |
| 2000 | 20348.26 | 20312.2 | Surgical | 1895 | 1270 | Surgical | 0 |
| 3000 | 31493.03 | 31342.44 | Surgical | 3445 | 2570 | Surgical | 1000 |
| 4000 | 42637.8 | 42372.68 | Surgical | 4995 | 3870 | Surgical | 2000 |
| 5000 | 53782.56 | 53402.92 | Surgical | 6545 | 5170 | Surgical | 3000 |

*Table 2: Optimal choice according to expected value and rank dependent valuations conditional on threshold with calculations for compensating threshold ε.*

Declarations:

Funding: Not Applicable

Conflicts of interest: The author has no conflict of interest to declare.

Availability of data and materials: Code used for computing rank dependent utility in the appendix is available upon request.

Ethics approval: Not Applicable

Author contributions: This paper is written by a single author

Informed consent: Not Applicable